\documentclass[aps,pra,reprint,floatfix,amsmath,amssymb,longbibliography]{revtex4-2}
\usepackage[utf8]{inputenc}
\usepackage{graphicx}
\usepackage{bm}
\usepackage{physics}
\usepackage{float}
\usepackage{hyperref}
\usepackage{color}
\usepackage[dvipsnames]{xcolor}
\usepackage{todonotes}
\hypersetup{
    colorlinks,
    linkcolor={red!50!black},
    citecolor={blue!50!black},
    urlcolor={blue!80!black}
}
\begin{document}

\title{\textcolor{black}{High fidelity control of a many-body Tonks--Girardeau gas with an effective mean-field approach}}
\author{Muhammad S.~Hasan$^1$}
\author{T.~Fogarty$^1$}
\author{J.~Li$^2$}
\author{A.~Ruschhaupt$^2$}
\author{Th.~Busch$^1$}
\address{$^1$Quantum Systems Unit, Okinawa Institute of Science and Technology Graduate University, Okinawa, Japan}
\address{$^2$School of Physics, University College Cork, Cork, Ireland }

\date{\today}

\begin{abstract}
    Shortcuts to adiabaticity (STA) are powerful tools that can be used to control quantum systems with high fidelity. They work particularly well for single particle and non-interacting systems which can be described exactly and which possess invariant or self-similar dynamics. However, finding an exact STA for strongly correlated many-body systems \textcolor{black}{can be} difficult, as their complex dynamics \textcolor{black}{may not} be easily described, especially for larger systems \textcolor{black}{that do not possess self-similar solutions}. Here, we design STAs for \textcolor{black}{one-dimensional bosonic gas in the Tonks--Girardeau limit by} using a mean-field approach that succinctly captures the strong interaction effects through a \textcolor{black}{quintic nonlinear term in the Schr\"odinger equation}. We show that for the case of the harmonic oscillator with a time-dependent trap frequency the mean-field approach works exactly and recovers the well-known STA from literature. To highlight the robustness of our approach we also show that it works effectively for anharmonic potentials, achieving higher fidelities than other typical control techniques.
\end{abstract}

\maketitle
\section{Introduction}
\label{intro}
The experimental control available to prepare, manipulate and measure ultra-cold atomic systems has pushed these systems to the forefront of being simulators for quantum many-body physics \cite{Bloch_Review:2008, Gross:2017}. Since their first realisation in 1995, atomic Bose-Einstein condensates (BECs) have become one of the best testbeds available to explore and scrutinise many aspects of quantum physics. The precise control over larger numbers of degrees of freedom have allowed the investigation of quantum phase transitions, the simulation of condensed matter systems, the study of many-body quantum interference and the control over atom-photon interactions, among many others \cite{Schafer:2020, Takahashi:2022}.

The main difficulty in developing full control over the dynamics of interacting quantum many-body systems is that their high spectral density makes it hard to experimentally isolate specific states. To do this with high fidelity, one usually needs to resort to  techniques that rely on adiabatic evolution and which therefore do not provide the system enough energy to make unwanted transitions. However, this leads to additional challenges in real-time applications since the time consuming nature of adiabatic evolution can lead to additional noise and decoherence in the system.  This intersection between the need for practicality and the desire for high fidelity gave rise to the area of shortcuts to adiabaticity \textcolor{black}{(STA)} \cite{Muga:2009, XiChen:2010}. STA techniques manipulate the evolution of a physical system by driving the dynamics in the direction of the target state in a rapid non-adiabatic manner, while still achieving high fidelities. They therefore provide an alternative to the time-consuming adiabatic methods. Since their inception in 2009, STAs have found countless applications in areas where fast and robust quantum control is required \cite{Andreas:2012}, including quantum thermodynamics, atom transport and state preparation  \cite{torrontegui:2013, STA_review:2019, cardenas2023shortcuts}. The field is constantly evolving and recently enhanced STA have been introduced and applied \cite{Chris:2020, Vladimir:2022, Manuel:2023}. 

STAs have mostly been developed for single particle and mean-field systems where scale invariance can be exploited, and experiments involving BECs and ultracold Fermi gases in harmonic traps have shown that STAs are technically feasible using modern techniques \cite{Schaff:2011, Rohringer:2015, Deng:2018, Diao:2018}. However, outside of these regimes practical STA protocols are rare, as controlling the increased number of degrees of freedom in interacting many-body systems can be a complex and difficult endeavour. To design experimentally realisable control ramps for such systems one therefore has to resort to approximate STA techniques, which can give high fidelity control via local potentials over a wide range of interactions \cite{Polkovnikov:2017,Li:18, Thomas:2019, Alan:2019, Huang2021}. Developing suitable STA techniques for many-body systems is therefore of significant interest, particularly in the strongly interacting regimes where large classical and quantum correlations are present. 

In this work we focus on the efficient control of a specific quantum many-body system consisting of strongly interacting bosons in the Tonks--Girardeau (TG) limit \cite{Girardeau:1960, kinoshita:2006}. The gas is \textcolor{black}{confined to a one-dimensional geometry in a time-dependent trapping potential, whose frequency is dynamically changed to realize a fast compression of the system without creating unwanted excitations.} While such a system requires a full quantum mechanical many-body treatment, we use a mean-field model to describe some aspects of its dynamics effectively \cite{Kolomeisky:2000}. We show that this approach allows us to efficiently calculate  STA pulses for systems with large particle numbers and strong interactions, and characterise its effectiveness compared to other commonly used control ramps through the many-body fidelity. Moreover, in the case of a harmonic trapping potential we retrieve the well known (and exact) STA pulse for non-interacting particles \cite{XiChen:2010}, while for anharmonic traps our method exhibits high fidelity control and robustness. The manuscript is organised as follows: in Section \ref{Sec:TG_KL} we outline the model of the TG gas and its mean-field description. In Section \ref{sec_STA} we derive STAs using the mean-field solution for harmonic and anharmonic traps, and discuss their results for different control parameters. Finally, in Section \ref{Sec:conclusion} we discuss potential applications for our approach and conclude. 

\section{System}
\label{Sec:TG_KL}
The quantum many-body system we consider is an ultracold gas of neutral bosonic atoms of mass $m$, which is strongly confined in two spatial directions such that it can be effectively described as a one-dimensional system. In the remaining direction the atoms are trapped in a potential $V(x,t)$, which can be modulated in time.  Since the scattering processes at low temperatures can be described by point-like interactions of strength $g$~\cite{Pethick:2008}, the many-body Schr\"odinger equation can be written as
\begin{align}
    i\hbar\frac{\partial}{\partial t}\Phi(\mathbf{x},t)=\left[ -\sum_{j=1}^N\frac{\hbar^2}{2m}\right.& \frac{\partial^2 }{\partial x_j^2} + V(x_j,t)\nonumber\\
    &\left. + g \sum_{k>j}\delta(x_k-x_j) \right]\Phi(\mathbf{x},t)\;.
    \label{eq:MB_TG}
\end{align}
Here $\mathbf{x}=\{x_1,x_2,\dots,x_N \}$ are the coordinates of the $N$ bosons. In the TG limit of infinite repulsive interactions, $g\rightarrow\infty$, the system can be famously mapped onto a system of non-interacting fermions, for which the many-body wavefunction can be written in terms of single-particle states as $\Phi_F(\mathbf{x},t)=\det[\varphi_n(x_j,t)]$ \cite{Girardeau:1960}. From this the bosonic many-body wavefunction can be obtained via $\Phi(\mathbf{x},t)=A(\mathbf{x})\Phi_F(\mathbf{x},t)$, where $A(\mathbf{x})= \prod_{k>j} \text{sgn}(x_k - x_j)$
is the unit anti-symmetric function. The description of the dynamics of the system can therefore be reduced to $N$ single particle Schr\"odinger equations of the form
\begin{equation}
    i\hbar \frac{\partial \varphi_j}{\partial t} = \left[ -\frac{\hbar^2}{2m} \frac{\partial^2 }{\partial x_j^2} + V(x_j,t) \right]\varphi_j\,.
    \label{single_part_equation}
\end{equation}
Since this mapping from a strongly interacting gas to a collection of non-interacting single particles often makes the simulation of the dynamical processes easier, it also seems an enticing approach to construct shortcuts to adiabaticity for this many-body system, since many methods for constructing STAs for single particle states exist \cite{Hatomura:2023}. However, for a many-body Fermi gas in an anharmonic potential each particle sees a different local spectral density and therefore they need to be optimised separately \cite{Polkovnikov:2017, TianNiu:2020}. If one requires that the STA needs to be driven by a change in a global potential, one has to either average over all single particle STAs \cite{Thomas:2020} or pick an optimal individual STA \cite{TianNiu:2020}. While both can be effective, neither of these is perfect for the whole system.

\begin{figure}[tb]
    \includegraphics[width=\linewidth]{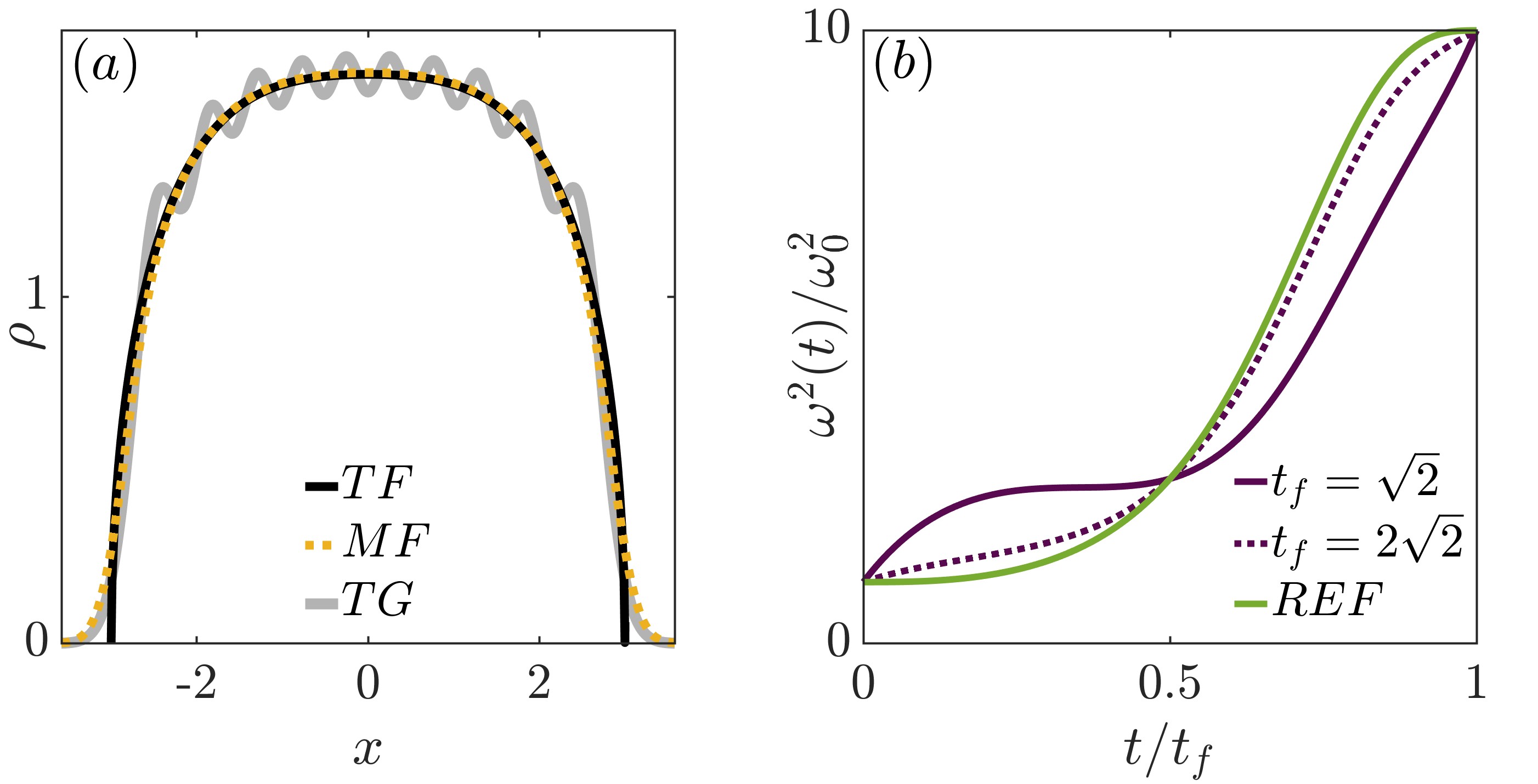}
	\caption{$(a)$ Densities from the exact solution of the non-linear Eq.~\eqref{eq:kolomeisky} (dotted yellow, MF),  for the corresponding TG gas from Eq.~\eqref{single_part_equation} (grey, TG) and from the Thomas--Fermi approximation to  Eq.~\eqref{eq:kolomeisky} (black, TF). All are for $N=10$ atoms trapped in a harmonic potential (see Sec.~\ref{sec_STA} for scaling).  The density of the TG gas displays Friedel oscillations, but otherwise aligns well with the density obtained from the MF Eq.~\eqref{eq:kolomeisky}. $(b)$ The shape of the ramp in Eq.~\eqref{HO_ramp} as a function of the scaled time for two overall times (purple, solid and dotted) and the reference case ($\ddot{b}=0$ in Eq.\eqref{HO_ramp}, green, REF) \textcolor{black}{for a trap compression of $\omega^2_f = 10 \omega_0^2$.}}
	\label{TG_KL_density}
\end{figure}

In this work we propose a different approach by approximating the exact many-body TG system using a non-linear mean-field (MF) description for the gas wave-function. It has been shown that the density distribution of the TG gas can be well approximated using a \textcolor{black}{quintic} mean-field equation of the form \cite{Kolomeisky:2000}
\begin{equation}
    i\hbar \frac{\partial \psi}{\partial t} = \bigg[ -\frac{\hbar^2}{2m} \frac{\partial^2 }{\partial x^2} + V(x,t) + \frac{\hbar^2\pi^2}{2m}|\psi|^4 \bigg]\psi\,,
\label{eq:kolomeisky}
\end{equation}
with $N=\int|\psi|^2 dx$. Here the strong interactions between the bosonic atoms are described by the \textcolor{black}{quintic} non-linear term, and the resulting density distribution $\rho_{MF}=|\psi|^2$ closely matches the one of a corresponding TG gas, $\rho_{TG}=\sum_{j=1}^{N} |\varphi_j|^2$ (see Fig.~\ref{TG_KL_density}$(a)$). In the following, we will derive a shortcut for the non-linear system given by Eq.~\eqref{eq:kolomeisky} and demonstrate its applicability to the many-body TG gas described by Eq.~\eqref{eq:MB_TG}. Except for a class of scale invariant systems \cite{Muga:2009, Tim:2020, Adolfo:2020, Adolfo:2021, Yang:2022},
it is difficult to find exact STAs for nonlinear systems. However, effective approximate methods have been devised \cite{XiChen:2010, XiChen:2011, Torrontegui:2012} and successfully experimentally implemented \cite{Masuda:2014, Dupont-Nivet:2018, Corgier:2018}. Our STA will in general also be approximate, as the MF equation does not fully capture the intricacies of the full many-body state, exemplified by the lack of Friedel oscillations in Fig.~\ref{TG_KL_density}$(a)$. 
However, we will show that our approach is amenable to arbitrary trapping potentials, and can become exact in certain limits. 

\section{Design and implementation of shortcuts}
\label{sec_STA}
\subsection{Harmonic oscillator}
In the following we will focus on power-law potentials of the form \textcolor{black}{$V(x,t)= \frac{1}{2}{m\omega^2(t)}(x^2 + \gamma x^4)$}, where $\gamma$ is the anharmonicity factor and \textcolor{black}{$\omega(t)$} corresponds to a time-dependent trapping \textcolor{black}{frequency and $m$ is the mass of the particles. Our initial state is the groundstate of the trapping potential with frequency $\omega(0)=\omega_0$ and we want to design $\omega(t)$ such that at the end of the driving the state is in the groundstate of a trapping potential with frequency $\omega(t_f)=\omega_f$}. Here we scale all the energies in terms of the harmonic oscillator energy \textcolor{black}{$\hbar\omega_0$} and all lengths in terms of \textcolor{black}{$a_0 = \sqrt{\frac{\hbar}{m\omega_0}}$, where $\omega_0$ is the trapping frequency at $t=0$}. This leads to the anharmonicity factor $\gamma$ being scaled in units of $a_0^2$ and the \textcolor{black}{time dependent trapping frequency in terms of $\omega_0$}. First we investigate the harmonic oscillator ($\gamma=0$) which is oft-studied in the STA literature due to its analytical tractability and experimental relevance \cite{Torrontegui:2011, XiChen:2011, Ibanez:2011, Torrontegui:2012, Abah:2018, Modugno2020, Huang2021}. To derive the STA for Eq.~\eqref{eq:kolomeisky} we adopt a scaling ansatz for the wavefunction of the form \cite{Muga:2009}
\begin{equation}
\psi(x,t) = \frac{1}{\sqrt{b(t)}}e^{-i\beta(t)}e^{-i\alpha(t) x^2} \phi\left(\frac{x}{b(t)},t\right),
\label{ansatz}
\end{equation}
where $\alpha(t)$ is a chirp, $\beta(t)$ is a global phase and \textcolor{black}{$b(t)\in \mathbb{R}$} is the dynamical scaling parameter describing the width of the state \cite{Castin:1996, Perez:1997, Abdullaev:2001}. Inserting this ansatz into the mean-field  Eq.~\eqref{eq:kolomeisky} and making the scaling transformation $y=\frac{x}{b(t)}$, 
we get
\begin{multline}
\mathcolor{black}{i \frac{\partial \phi}{\partial t}= 
-\frac{1}{2b^2} \frac{\partial^2 \phi}{\partial y^2} + \mathcolor{black}{\left[\frac{1}{2}\omega^2(t) - \dot{\alpha} + 2\alpha^2 \right]} y^2 b^2 \phi + \frac{\pi^2}{2b^2} 
|\phi|^4 \phi} \\
\mathcolor{black}{+ \left[ \frac{i \dot{b}}{2b} - \dot{\beta} + i \alpha \right]\phi  + \left[ 2 i \alpha + \frac{i\dot{b}}{b} \right] y \frac{\partial \phi}{\partial y}},
\label{aux_equation}
\end{multline}
where the dots represent the time derivative. We set the chirp $\alpha = -\frac{1}{2}\frac{\dot{b}}{b}$ so that the last term in Eq.~\eqref{aux_equation} vanishes. We then demand that \textcolor{black}{$\frac{1}{2}\left[\omega^2(t) -\dot{\alpha} + 2\alpha^2 \right] = \frac{\omega_0^2}{2b^4}$} so that the equation resembles a harmonic oscillator with stationary trapping \textcolor{black}{frequency $\omega_0$}. Substituting the expression for $\alpha$ gives the equation that establishes the relation between $b$ and the trapping \textcolor{black}{frequency $\omega(t)$} as 
\begin{equation}
    \mathcolor{black}{\omega^2(t) = \frac{\omega_0^2}{b^4}-\frac{\ddot{b}}{b}\;}.
    \label{HO_ramp}
\end{equation}
If we substitute \textcolor{black}{$\omega(t)$} into Eq.~\eqref{aux_equation} we get
\begin{equation}
i \frac{\partial \phi}{\partial t}= 
-\frac{1}{2b^2} \frac{\partial^2 \phi}{\partial y^2} +\mathcolor{black}{\frac{\omega_0^2}{2b^4}} y^2 b^2 \phi + \frac{\pi^2}{2b^2}|\phi|^4 \phi\;.
\label{reduced_aux_eqn}
\end{equation}
Performing a transformation in $t$ such that $\tau = \int_0^t \frac{dt}{b^2}$ results now in the rescaled Schr\"odinger equation for a harmonic oscillator with stationary trapping \textcolor{black}{frequency}
\begin{equation}
i \frac{\partial \Psi}{\partial \tau} = \left[ -\frac{1}{2} \frac{\partial^2 }{\partial y^2} +\mathcolor{black}{\frac{1}{2}\omega_0^2} y^2  +  \frac{\pi^2}{2}|\Psi|^4 \right]  \Psi
\end{equation}
where $\Psi(y,\tau) = \phi(y,t)$. 

It is worth noting that Eq.~\eqref{HO_ramp} is the well-known Ermakov equation of motion which is synonymous with describing the integrable dynamics of the harmonic oscillator \cite{Muga:2009}. The Ermakov equation also appears in single particle and many-body scale invariant systems, of which the non-interacting Fermi gas and TG gas are examples \cite{Kagan:1996, Deffner:2014, Yang:2022}. It is also interesting to note that in the case of changing the trapping strength of a harmonic oscillator in one dimension within the non-linear framework of the Gross-Pitaevskii equation, where the nonlinearity is quadratic, $g|\psi|^2$, the ansatz \eqref{ansatz} works only approximately, unless the interaction strength is rescaled in time by $b(t)$ \cite{Muga:2009}. \textcolor{black}{However, in the case of a quintic non-linearity, the ansatz \eqref{ansatz} works exactly, which should translate into the ability of Eq.~\eqref{eq:kolomeisky} to exactly describe the non-equilibrium dynamics of many-body quantum states in the strongly interacting limit as first shown in Ref.~\cite{kolomeisky:2001} for a harmonic trap
of fixed frequency}. 

To find the exact form of \textcolor{black}{$\omega(t)$ by inverse engineering} requires one to choose a suitable ansatz for the scaling factor $b(t)$. A common and convenient choice is a polynomial of the form $b(t) = \Sigma_{i=0}^5 a_i t^i$, with the boundary conditions at the beginning ($t=0$) and at the end of the STA ($t=t_f$) given by $b(0) = 1$, \textcolor{black}{$b(t_f) = (\frac{\omega_0}{\omega_f})^{\frac{1}{2}}$} and $\dot{b}(0) = \dot{b}(t_f) =  \ddot{b}(0) = \ddot{b}(t_f) = 0$. The shape of the resulting STA is strongly dependent on the ramp duration $t_f$ (see Fig.~\ref{TG_KL_density}$(b)$), with larger modulations of the trap strength required to drive the system to the target state for $t_f$ smaller than the characteristic timescale of the trap \textcolor{black}{$1/\omega_f$}. Meanwhile, when the system is driven slowly, such that \textcolor{black}{$t_f\ggg 1/\omega_f$}, the STA ramp approaches the adiabatic limit defined by $\ddot{b}=0$, in which it is simply described by \textcolor{black}{$\omega^2(t)=\omega_0^2/b^4$}. In what follows we will use this as a convenient reference ramp (REF) to compare to the STAs derived for arbitrary $t_f$.

To confirm the effectiveness of the STA to drive quantum many-body dynamics we apply it to the many-body TG state described by Eq.~\eqref{eq:MB_TG} and its effective mean-field description in Eq.~\eqref{eq:kolomeisky}. To quantify the success of the driving protocol we calculate the overlap between the density of the state at the end of the shortcut, $\rho(x,t_f)$, and that of the target state, $\rho^T(x)$, as
\begin{equation}
    \mathcal{O}(t_f)=\bigg|\int_{-\infty}^{\infty}\sqrt{\rho(x,t_f)\rho^T(x)}dx\bigg|^2\,,
\end{equation}
which takes the value of $\mathcal{O}(t_f)=1$ when the target state is reached exactly. 

\begin{figure}[tb]
    \includegraphics[width=\linewidth]{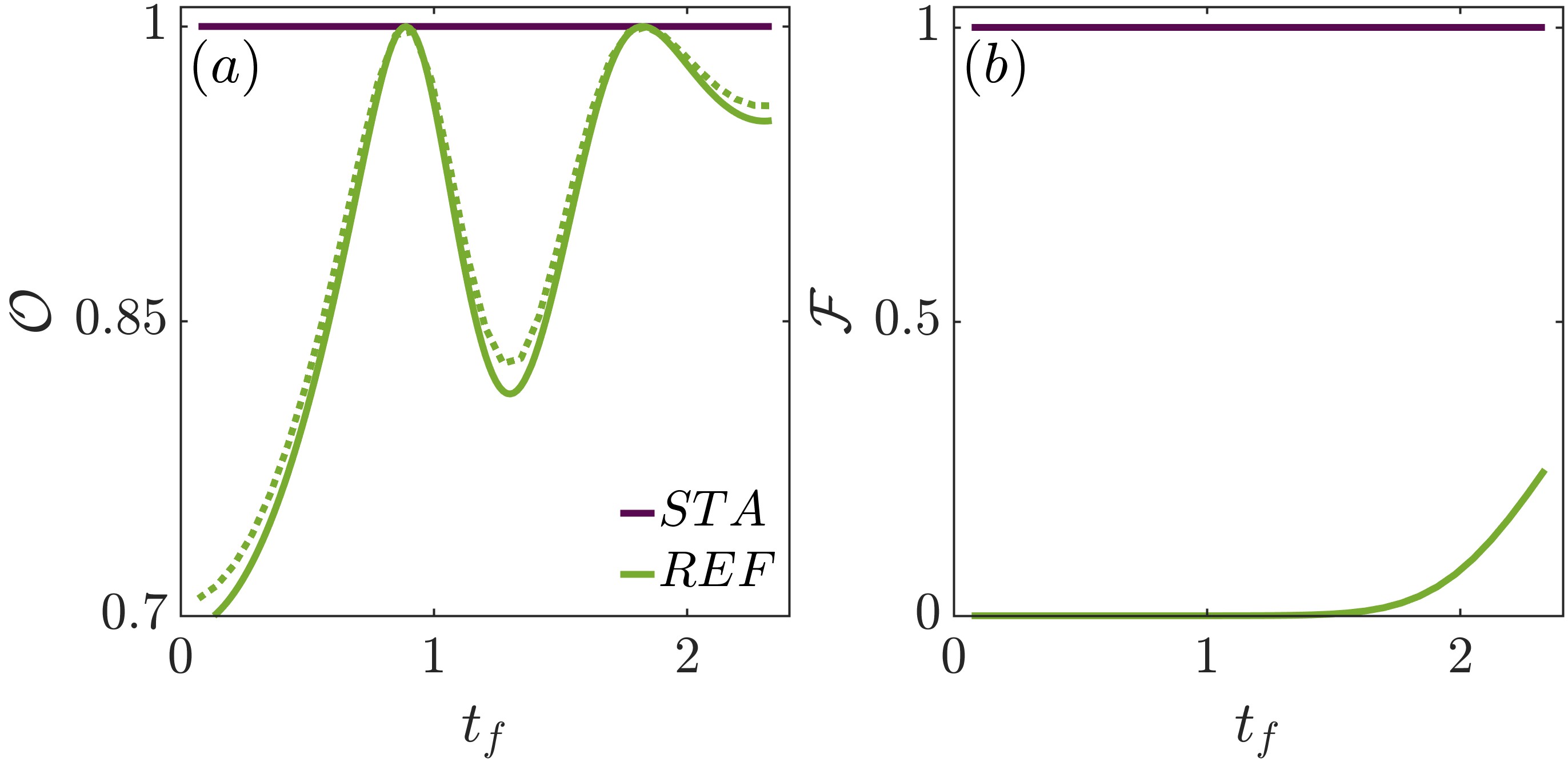}
	\caption{$(a)$ Density overlap $\mathcal{O}$ of the TG gas (solid) and obtained from the mean-field approach (dotted) as a function of final time $t_f$ for compressing a harmonic trap ($\gamma=0$) with \textcolor{black}{$\omega^2_f = 10\omega^2_0$} using the STA (Eq.~\eqref{HO_ramp}, purple) and reference ramps ($\ddot{b}(t)=0$ in Eq.~\eqref{HO_ramp}, green), for $N=10$ particles. $(b)$ Many-body fidelity $\mathcal{F}$ as a function of $t_f$ for the same parameters.}
	\label{HO_fid}
\end{figure}

The density overlap is shown in Fig.~\ref{HO_fid}$(a)$, where one can see that values of exactly $\mathcal{O}(t_f)=1$ can be obtained for both the TG gas and its mean-field description, confirming that the STA derived above works perfectly for both the nonlinear and many-body systems. In comparison, using the reference ramp gives density overlaps smaller than unity, which shows that the target state is not reached (except at the so-called \textit{magic times}, where the final state is dynamically unstable). We also note that the density overlap of the mean-field description closely matches that of the TG gas even when using the reference ramp, and only shows slight deviations when both systems are driven far from equilibrium, i.e.~when $\mathcal{O}(t_f)\ll 1$. This highlights the ability of the mean-field model to accurately describe the many-body density dynamics.

While the density overlap is a useful metric to compare the mean-field and TG dynamics, deeper insights can be obtained by considering the many-body fidelity which describes the overlap of the full many-body states
\begin{equation}
\mathcal{F}(t_f)=\bigg|\int_{-\infty}^{\infty}\Phi(\textbf{x},t_f)\Phi^T(\textbf{x})dx\bigg|^2\,.
\label{eq:MB_fid}
\end{equation}
This will, in particular, give information about the appearance of the orthogonality catastrophe, which can be detrimental to quantum control of large systems \cite{Thomas:2020}. 
\textcolor{black}{The quantity in Eq.~\eqref{eq:MB_fid} can only be calculated for the many-body TG model and can be conveniently computed by taking advantage of the Fermi-Bose mapping theorem and the decomposition of the many-body dynamics in terms of single particle states as}
\begin{equation}
    \mathcal{F}(t_f) = \left|\mathbf{det}[\mathcal{A}(t_f)]\right|^2\;,
    \label{many_body_fidelity}
\end{equation}
where the matrix elements $ \mathcal{A}_{l,m}(t_f)$ are given by
\begin{equation}
    \mathcal{A}_{l,m}(t_f) = \int_{-\infty}^{\infty} \varphi_{l}^*(x,t_f) \varphi^T_{m}(x)\; dx,
\end{equation}
with $\varphi_l(x,t_f)$ being the single particle states at the end of the STA and $\varphi^T_{m}(x)$ the target single particle states.

The many-body fidelity for the STA and the reference ramp is shown in Fig.~\ref{HO_fid}$(b)$ as a function of $t_f$. Echoing the density overlap we find that $\mathcal{F}(t_f)=1$ for all ramp times when using the STA, as it perfectly drives each of the $N$ single particle states to their respective target states without any loss in fidelity. Again, this is to be expected as the STA ramp for the harmonic oscillator is exact and independent of the number of particles \cite{adolfo:2011, TianNiu:2020, Adolfo:2021}.
In comparison, the many-body fidelity of the reference ramp shows that the final state is orthogonal to the target state for ramp times $t_f<1$ which is in stark contrast to the density overlap in Fig.~\ref{HO_fid}$(a)$. This highlights that care must be taken when quantifying the adiabaticity of driven quantum systems by considering the density alone, especially for many-body states. In what follows we will therefore only focus on the many-body fidelity to quantify the effectiveness of the STA protocols.  
\subsection{Anharmonic traps}

Next we explore the effects of an anharmonicity, $\gamma> 0$, in the trapping potential. In general to derive an STA for anharmonic traps the scaling approach used previously can not be employed. This is due to the breaking of many-body scale invariance in the system which as a consequence requires individual equations of motion for each single particle state $\varphi_n(x,t)$ \cite{TianNiu:2020}. This makes it exceedingly difficult to find an exact STA using local potentials only.  We are therefore restricted to using approximate STA techniques to control the system with high fidelity, such as variational approaches \cite{Perez:1996, Perez:1997, Jing:2016}. These STAs rely on minimising the action on the quantum state by solving the Euler-Lagrange equations. Like in the scaling approach we will design the STA based on the mean-field approximation for the TG gas in Eq.~\eqref{eq:kolomeisky}, which has the following Lagrangian density
\begin{align}
    \mathcal{L} = \int_{-\infty}^{\infty} dx \Bigg[& \frac{i}{2}(\psi\dot{\psi}^* - \psi^*\dot{\psi}) + \frac{1}{2}\left|\frac{d}{dx}\psi\right|^2 \nonumber\\ 
    &+  \mathcolor{black}{\frac{1}{2}\omega^2(t)}(x^2 + \gamma x^4) |\psi|^2 + \frac{\pi^2}{6} |\psi|^6 \Bigg]\;,
    \label{general_lag}
\end{align}
with the last term describing the interaction. The variational approach is strongly dependent on the ansatz for the time evolved state $\psi(x,t)$, and inspired by its success, we choose the same form as for the scaling approach given in Eq.~\eqref{ansatz} and which again depends on the scaling parameter $b(t)$. Applying a change of variables $x = y b(t)$ and dropping the explicit notation of the time-dependence of the Lagrangian parameters, the Lagrangian itself then becomes
\begin{align}
    \mathcal{L} = -\dot{\beta}N - b^2\dot{\alpha}W 
    +&\frac{1}{2} \left[ \frac{F}{b^2} +4\alpha^2b^2W \right]+\mathcolor{black}{\frac{1}{2}\omega^2(t)}b^2W \nonumber\\ 
    &+ \mathcolor{black}{\frac{1}{2}\omega^2(t)} \gamma b^4 J+ \frac{\pi^2}{6b^2}K\;,
\label{eq:L}
\end{align}
where $N = \int_{-\infty}^{\infty} dy~ |\phi(y)|^2$, $W = \int_{-\infty}^{\infty} dy ~y^2 |\phi(y)|^2$, $F = \int_{-\infty}^{\infty} dy~ \big(\frac{d \phi(y)}{dy} \big)^2$, $J = \int_{-\infty}^{\infty} dy~ y^{4} |\phi(y)|^2$ and $K = \int_{-\infty}^{\infty} dy ~|\phi(y)|^6$.
The Euler-Lagrange equations 
\begin{equation}
    \frac{\partial \mathcal{L}}{\partial s}-\frac{d}{dt}\bigg(\frac{\partial \mathcal{L}}{\partial \dot{s}}\bigg)=0\;,
    \label{eq:EL}
\end{equation}
for $s = \alpha, b$ and $\beta$ can then be straightforwardly calculated and for the variational parameter describing the phase, $\beta$, we simply obtain particle number conservation. The other two give 
\begin{align}
    4\alpha^2b - 2\dot{\alpha}b - \ddot{b} &= 0, \\
    4bW\left( \alpha^2-\frac{\dot{\alpha}}{2} \right)-\frac{1}{b^3}\left(F+\frac{\pi^2 K}{3}\right)
    \nonumber\\ 
    +b \mathcolor{black}{\omega^2(t)}(W +2\gamma b^2J) &= 0. 
\label{eq:EL3}
\end{align}
Solving these for $\omega^2(t)$ gives the expression for the ramp as
\begin{equation}
    \mathcolor{black}{\omega^2(t) = \frac{1}{b^4[W+2b^2\gamma J]}\left[ \left(F +\frac{\pi^2}{3}K \right) - \ddot{b}b^3 W\right]\;},
    \label{lambda_exp}
\end{equation}
where $W$, $F$, $J$, and $K$ are obtained by integrating the ansatz wavefunction $\phi(y)$. The boundary conditions on the scaling factor $b$ is chosen to be same as before except for $b(0)\equiv b_0$ and $b(t_f)\equiv b_f$ which are found by solving the polynomial equations
\begin{align}
    \mathcolor{black}{2\gamma J b_0^6 + Wb_0^4 - \frac{F}{\omega^2_0} - \frac{\pi^2 K}{3\omega^2_0}&=0}, \\
    \mathcolor{black}{2\gamma J b_f^6 + Wb_f^4 - \frac{F}{\omega^2_f} - \frac{\pi^2 K}{3\omega^2_f}&=0}.
\end{align}

The choice of the functional form of $\phi(y)$ is crucial for constructing an accurate STA and a common choice in harmonic and anharmonic traps is a Gaussian (G) ansatz $\phi_{\text{G}}(y) = \sqrt{N}\left( \frac{2}{\pi} \right)^{\frac{1}{4}}e^{-y^2}$. This is known to well approximate the density for weakly interacting Bose gases \cite{Schaff2:2011} and also to significantly simplify the calculations. However, it does not capture the broadening of the wavefunction due to the hardcore interactions in TG limit. To more accurately describe this a better choice is to use the Thomas--Fermi (TF) approximation of Eq.~\eqref{eq:kolomeisky}, \textcolor{black}{$\phi_{\text{TF}}(y) = \left( \frac{2\mu  - (y^2+ \gamma y^{4})}{\pi^2}\right)^{\frac{1}{4}}$}, where $\mu$ is the chemical potential. This choice is further motivated by the fact that Eq.~\eqref{lambda_exp} reduces to the harmonic oscillator STA Eq.~\eqref{HO_ramp} when $\gamma=0$. The integral constants in that case are then given by $W = \frac{N^2}{2\sqrt{2}}$, $K = \frac{3N^2}{\pi^2\sqrt{2}}$ and $F = -\frac{1}{2\sqrt{2}}$. In the limit $N\rightarrow\infty$ the $\frac{F}{W}$ term vanishes and the STA in Eq.~\eqref{HO_ramp} is recovered, showing that the variational approach becomes exact in the harmonic trap.  

\begin{figure}[t]
    \centering
    \includegraphics[width=\linewidth]{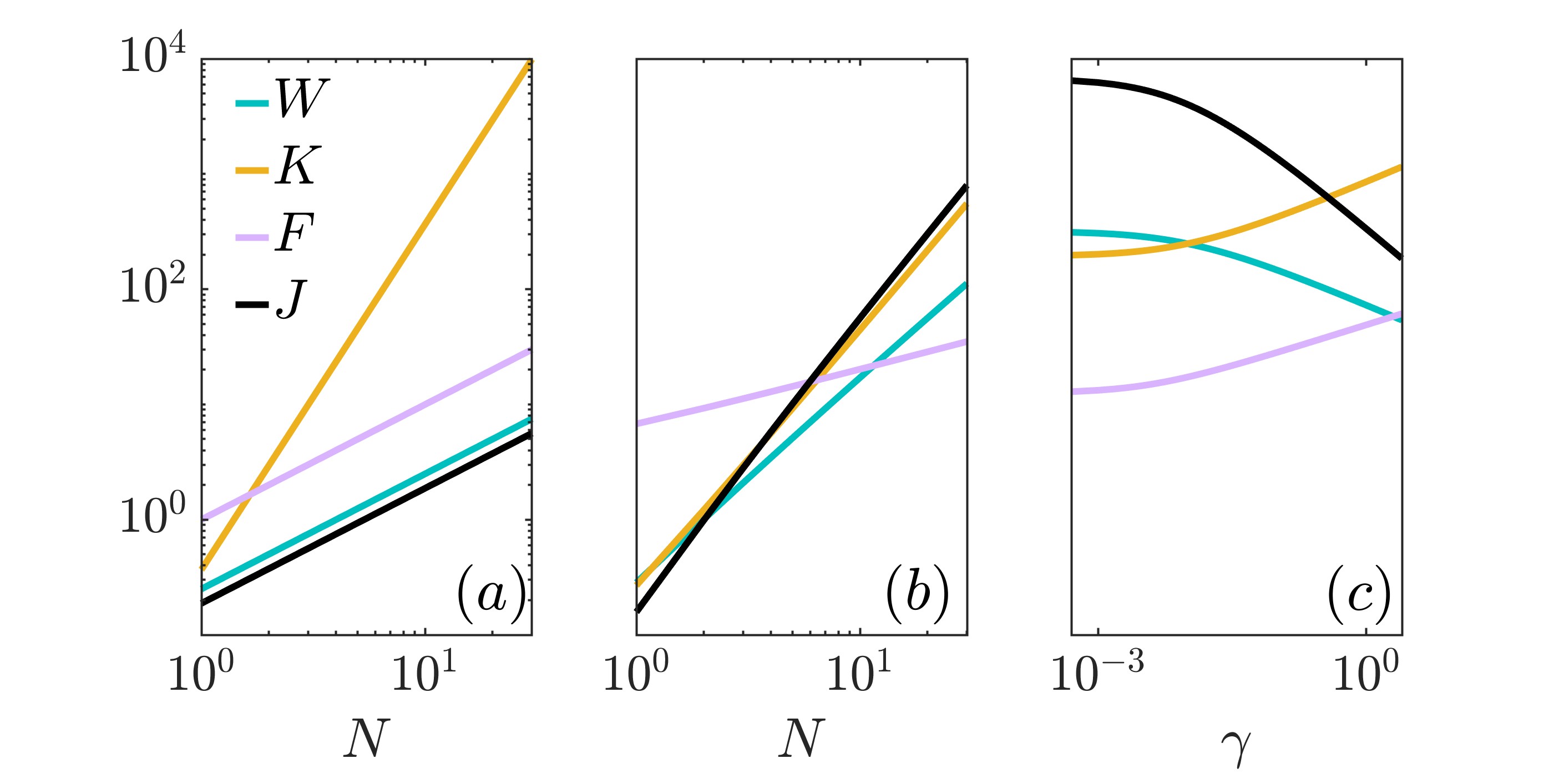}
    \caption{Integrals $W, K, F ~\text{and}~ J$ for anharmonicity factor \textcolor{black}{$\gamma=0.25$} as a function of $N$ for $(a)$ the Gaussian ansatz and $(b)$ the Thomas--Fermi ansatz. $(c)$ The integrals as a function of $\gamma$ for $N=30$ particles using the Thomas-Fermi ansatz. Note that the units for each curve are different and given in the text.}
    \label{fig:int_vs_N}
\end{figure}

In Fig.~\ref{fig:int_vs_N} we show the behaviour of the integral values appearing in the ramp expression \eqref{lambda_exp} as a function of $N$ for the Gaussian (panel $(a)$) and Thomas--Fermi ansatz (panel $(b)$), with a fixed anharmonicity strength \textcolor{black}{$\gamma=0.25$}. The presence of the anharmonic term does not affect the Gaussian ansatz and therefore the terms $W$, $F$ and $J$ are linear in $N$, while the contribution from the interaction term scales as $K\propto N^3$. On the contrary, the Thomas--Fermi ansatz is by construction strongly dependent on the anharmonicity of the trap, which in turn affects the scaling of these terms (see panels $(b)$ and $(c)$). It is immediately apparent that they differ from the harmonic oscillator case ($\gamma=0$) presented above, as $W\propto N^{1.745}$, $K\propto N^{2.285}$, $F \propto N^{0.49}$ and $J \propto N^{2.458}$ for an anharmonicity of \textcolor{black}{$\gamma=0.25$}. The resultant STA derived from the Thomas--Fermi ansatz is therefore more amenable to changes in the trap shape, which should allow for enhanced control of many-body dynamics in different trapping potentials.   

\begin{figure}[b]
    \includegraphics[width=\linewidth]{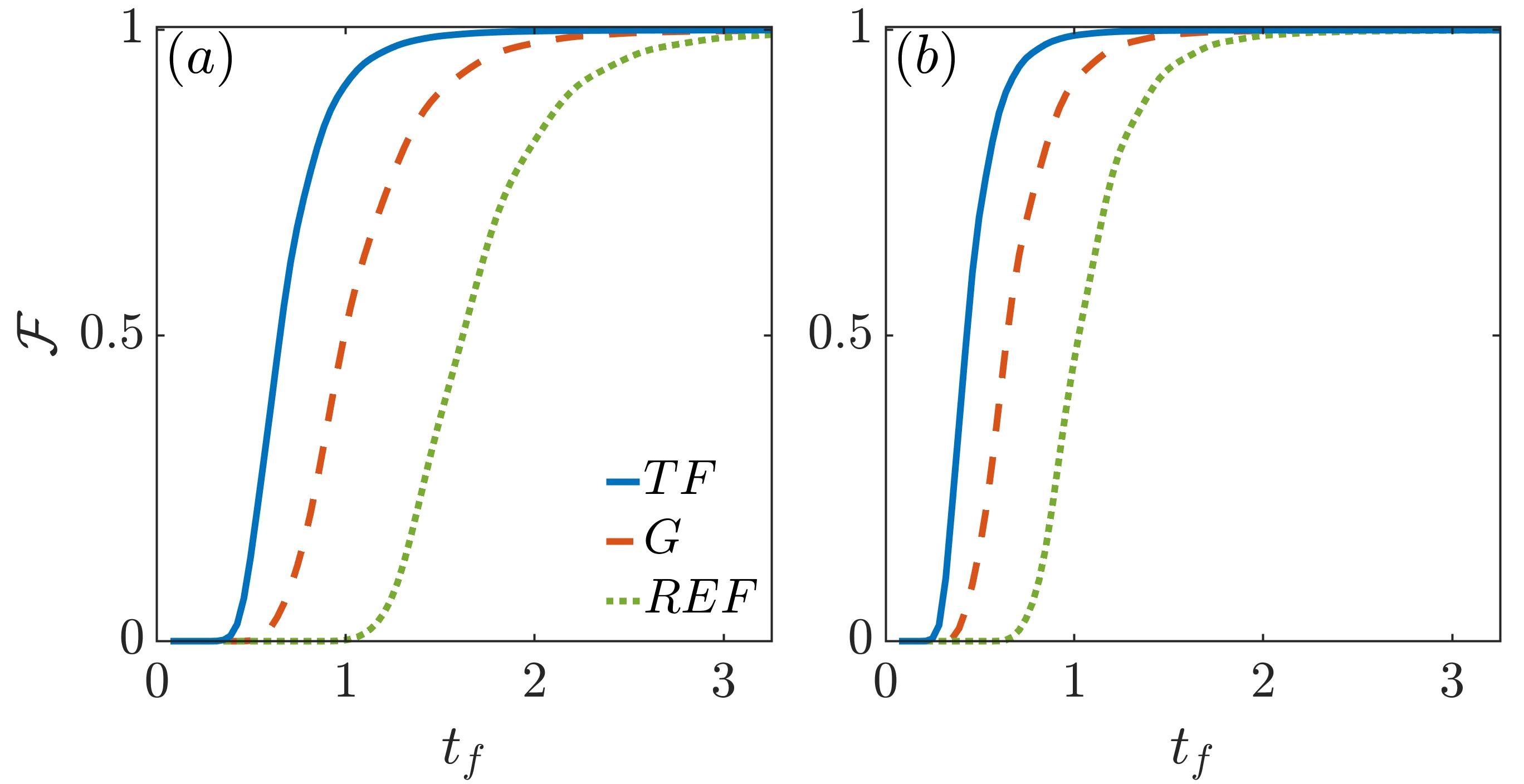}
	\caption{Many-body fidelity as a function of $t_f$ for ramps based on Thomas-Fermi (blue, solid, $TF$) and Gaussian (orange, dashed, $G$) ans\"atze, and the reference ramp (green, dotted, $REF$) for \textcolor{black}{$(a)~\gamma=0.25$ and $(b)~\gamma=1$} for fixed $N=30$ TG particles.}
	\label{mbo_vs_tf}
\end{figure}

The many-body fidelity as a function of $t_f$ for a trap compression of \textcolor{black}{ $\omega^2_f = 10\omega_0^2$} is shown in Fig.~\ref{mbo_vs_tf}. The results are shown for a system size of $N=30$, \textcolor{black}{for $\gamma=0.25$ in panel $(a)$ and $\gamma=1.0$} in panel $(b)$. Here we compare fidelities from STAs designed with both the Thomas--Fermi and Gaussian ans\"atze, and also the reference ramp. Both STAs perform well compared to the reference, which is not optimised for the chosen $t_f$. However, the STAs are not perfect as they rely on approximations to the full many-body state and their fidelities rapidly decrease for quick ramps $t_f<1$. Regardless, the Thomas--Fermi STA clearly outperforms the other ramps for all driving times $t_f$, suggesting it is an effective ansatz for the TG gas. 

\begin{figure}[t]
    \includegraphics[width=\linewidth]{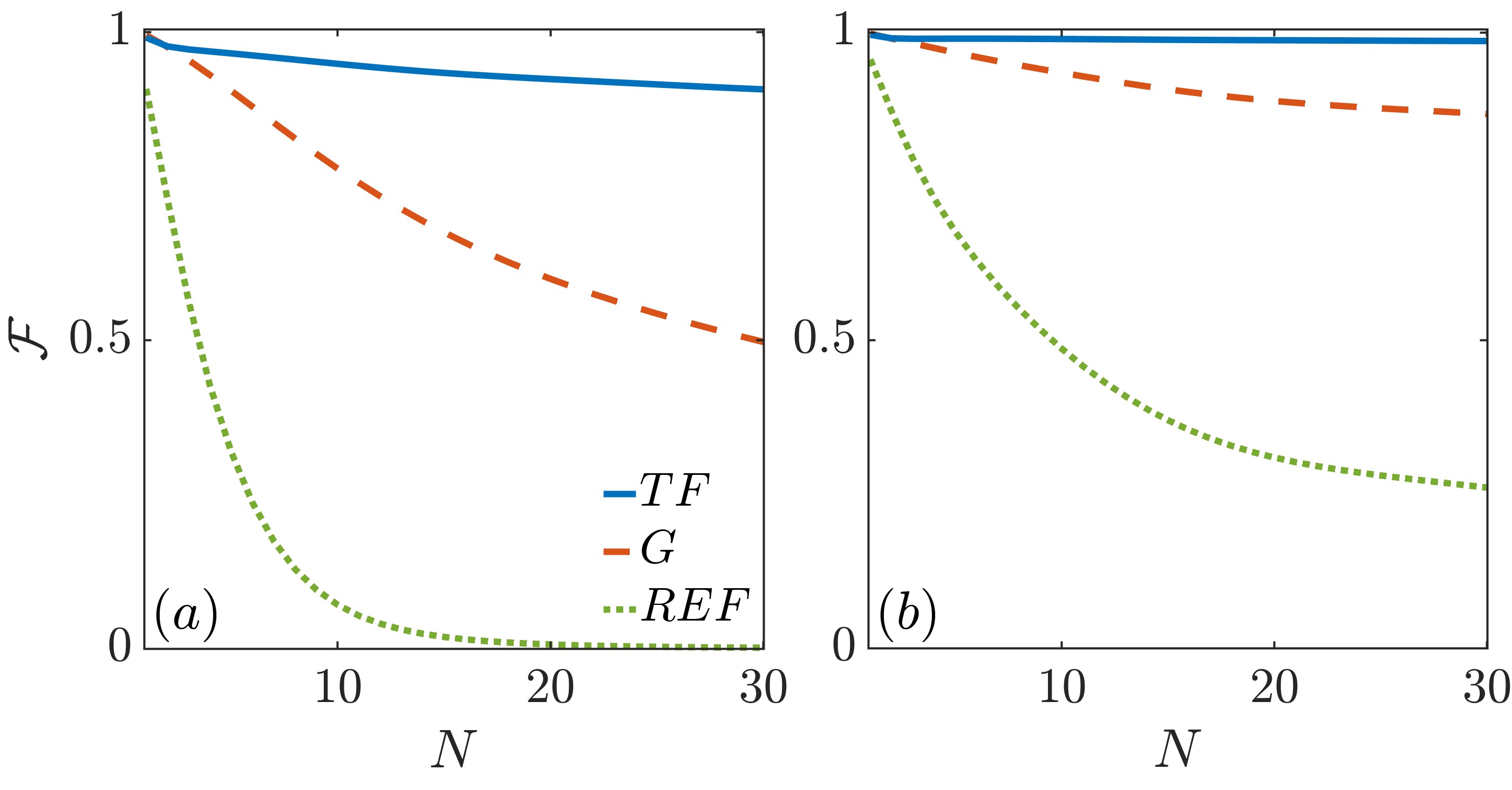}
	\caption{The many-body fidelity of the TG  gas as a function of $N$ for ramps based on the Thomas--Fermi (blue, solid, $TF$), Gaussian (orange, dashed, $G$) ans\"atze, and for the reference ramp (green, dotted, $REF$) at final time \textcolor{black}{$(a)~t_f=1$ and $(b)~t_f=\sqrt{2}$ for fixed anharmonicity $\gamma=0.25$}.}
	\label{mbo_vs_N}
\end{figure}

As mentioned previously, the many-body fidelity allows to explore the presence of orthogonality catastrophe (OC), whereby the overlap between two states decreases with increasing particle number $\langle \Psi| \Phi \rangle\sim N^{-\delta}$, with $\delta$ the strength of the perturbation \cite{Anderson:1967}. This has consequences for the ability to control large systems \cite{Polkovnikov:2017} as the quantum speed limit can vanish in the thermodynamic limit \cite{Bukov:2018, Thomas:2020}. While exact STAs allow to circumvent this, approximate STAs will nevertheless be affected by the OC as the variational ansatz is not exact and the effectiveness of the STAs will strongly depend on $N$. In Fig.~\ref{mbo_vs_N} we show the many-body fidelity as a function of $N$ for two ramps times, \textcolor{black}{$(a)$ $t_f=1$ and $(b)$ $t_f=\sqrt{2}$}. The OC is clearly visible in the decay of $\mathcal{F}$ when considering the dynamics of the reference ramp which vanishes for $N=30$ and \textcolor{black}{ $t_f=1$}. The approximate STAs perform better, arresting the decay of the fidelity and reducing the effects of the OC, especially when using the Thomas--Fermi ansatz. Indeed, while simplistic in description, the dependence of the Thomas--Fermi ansatz on both the trap shape and particle number, coupled with the mean-field approach presented in this work, suggest that it can be a powerful tool for controlling strongly interacting many--body states.
\section{Conclusion}
\label{Sec:conclusion}

We have shown that the \textcolor{black}{quintic} mean-field description of the TG gas can be used to design effective STAs for strongly interacting many-body states. Applying this mean-field approach to the harmonic oscillator potential shows that our chosen ansatz works exactly, resulting in the well-known STA ramp from literature. To test the robustness of our approach we have considered anharmonic potentials which require using a variational ansatz for the STA and which are therefore no longer exact. While perfect fidelity cannot be preserved for fast driving of the system, the use of the Thomas--Fermi approximation for the TG density in the variational ansatz allows to construct an efficient STA when compared to other trial wavefunctions. 

Computation of the STA using the mean-field description offers practical advantages over using the full many-body state, requiring less numerical resources and being applicable to any trapping potential where the Thomas--Fermi profile closely matches the systems density. This method could also be extended to consider weakly interacting many-body states, which can be approximated by the Gross--Pitaevskii equation, while intermediate interaction regimes could be described by a superposition ansatz combining both weakly and strongly interacting regimes ~\cite{Thomas:2019}. 

\section*{Acknowledgements}

This work was supported by the Okinawa Institute of Science and Technology Graduate University. The authors are grateful for the the Scientific Computing and Data Analysis (SCDA) section of the Research Support Division at OIST. M.S.H. thanks Prof. Nathan Harshman and Mr. David Rea for discussions. T.F. acknowledges support from JSPS KAKENHI Grant No. JP23K03290. T.F. and T.B. are also supported by JST Grant No. JPMJPF2221 and JSPS Bilateral Program No. JPJSBP120227414.
J.L. and A.R. acknowledge that this publication has emanated from research supported in part by a research grant from Science Foundation Ireland (SFI) under Grant Number 19/FFP/6951.  

\bibliography{references}
\end{document}